\documentclass[12pt]{article}
\usepackage[centertags]{amsmath}
\usepackage{graphicx}
\newcommand{\be}{\begin{equation}}
\newcommand{\ee}{\end{equation}}
\usepackage[centertags]{amsmath}
\usepackage{amsfonts}

\newcommand{\Ei}{E_{i}}

\newcommand{\eijk}{\epsilon_{ijk}}

\begin{document}
\title{\bf The dyon charge in noncommutative gauge theories}
\author{L.~Cieri and F.A.~Schaposnik\thanks{Associated with CICBA}
\\
{\normalsize\it Departamento de F\'\i sica, Universidad Nacional
de La Plata}\\ {\normalsize\it C.C. 67, 1900 La Plata, Argentina}
\\
{\normalsize \it CEFIMAS, Av.\,Santa Fe 1145,
 1069 Capital Federal,
Argentina} } \date{\hfill}

\maketitle
\begin{abstract}
We present an explicit classical dyon solution for the
noncommutative version of the Yang-Mills-Higgs model (in the
Prasad-Sommerfield limit) with a $\vartheta$-term. We show
 that the  relation  between classical electric and magnetic charges
also holds in noncommutative space. Extending the Noether approach to the case of
a noncommutative gauge theory, we analyze the effect of CP violation at the quantum level,
induced both by the $\vartheta$ term and by  noncommutativity
and we prove that the Witten effect formula for the dyon charge remains the same as in
ordinary space.
\end{abstract}
\section{Introduction}
Gauge theories coupled to Higgs scalars exhibit a remarkable
phenomenon, usually called Witten effect \cite{Witten}, related to
the $\vartheta$-angle. Indeed, if one adds to a Yang-Mills-Higgs
Lagrangian, a $\vartheta$-term,
\begin{equation}
\Delta L = \vartheta \frac{e^2}{32\pi^2}
\varepsilon^{\mu\nu\alpha\beta}tr \left(F_{\mu\nu}
F_{\alpha\beta}\right) \label{theta} \, ,
\end{equation}
which explicitly violates CP,  the electric charge $q_e$ of a dyon is
modified. Instead of being quantized,  as in the $\vartheta = 0$ case,
in units of the fundamental
charge $e$ -  as can be seen   using, for example,
 semiclassical quantization
arguments
 \cite{TW}- one has, for $\vartheta \ne 0
$
\begin{equation}
q_e =\left( n + \frac{\vartheta}{2\pi}\right)e \label{effect} \; ,
\;\;\; n \in Z
\end{equation}
This result corresponds to a Julia-Zee dyon \cite{JZ} with magnetic
charge $m=4\pi/e$. There are also arguments  leading to the
conjecture that other CP violating interactions may also induce a
shift of the dyon charge
 \cite{Witten}.

CP violation can be induced not just by adding new interactions to
the Yang-Mills-Higgs Lagrangian but by radically changing the
setting of the theory. This is the case of noncommutative gauge
theories (NCGT) where the introduction of noncommutation in
space-time coordinates has shown to affect the behavior under C, P
and T invariance \cite{Hayakawa}-\cite{Chaichian}. More
specifically, one can prove that when noncommutativity is restricted
to space coordinates,
\begin{eqnarray}
[x_i,x_j] &=& i \theta_{ij}  \, , \;\;\; i,j=1,2,3\nonumber
\\
~ [ x_i,x_0] &=& 0 \label{nc}
\end{eqnarray}
NCGT are not charge invariant. Only if the usual field
transformations are accompanied by a change of sign in
$\theta_{ij}$, charge invariance is recovered. Hence, if one takes
$\theta_{ij}$ as a fixed parameter which does not transform as
fields do, CP is violated, although CPT invariance is maintained
since parity invariance is not affected by the introduction of
$\theta$ and time reversal undergoes a   change that compensates
that in C.

It is then natural to pose the question whether  the dyon charge in
noncommutative gauge theories  receives a contribution from a CP
violating effect induced by noncommutativity even if the $\vartheta$
angle vanishes. Moreover, one could also ask how the addition of the
noncommutative version of the  term (\ref{theta}) modifies $Q$ when
both $\theta_{ij} \ne 0$ and $\vartheta \ne 0$.

 We
analyze these questions in the present paper and, to this end, we
first discuss the properties of the  dyon in a noncommutative
Yang-Mills-Higgs model with $U(2)$ gauge symmetry, calculate its
charge at the quantum level and also extend the theory in order to
include a $\vartheta$ term. Some of these issues were briefly
discussed in \cite{GN} starting from a monopole solution obtained
generalizing Nahm's equations that describes BPS solitons \cite{GN0}
(Some aspects of dyon solutions were also considered in
ref.{\cite{Bak}). Here, instead, we shall extend the more explicit
$U(2)$ monopole solution found in \cite{H1}-\cite{H2} to the case of
a dyon and then establish a noncommutative version of the Noether
theorem in order to define the operator ${\cal N}$ which generates
the $U(1)$ gauge transformations associated with electric charge. We
then discuss the issue of the Witten effect in noncommutative space.

\section{The Bogomolnyi bound and the noncommutative dyon equations}
The action for the noncommutative $U(2)$ Yang-Mills-Higgs system
that we consider is
\begin{equation}
S = {\rm tr} \int d^4 x \left(-\frac{1}{2} F_{\mu\nu} * F^{\mu\nu} +
 D_\mu \Phi * D^\mu\Phi \right) \label{uno}
\end{equation}
Gauge fields $A_\mu = A_\mu^At^A $ take values in the Lie algebra of
U(2) with generators $t^A, A=0,1,2,3$ ($t^0 = I/2, t^a = \sigma^a/2,
a=1,2,3$). $\Phi = \Phi^A t^A$ is the Higgs multiplet and we
consider the Prasad-Sommerfield limit \cite{PS} in which the
symmetry breaking potential vanishes. Covariant derivatives and
field strength are defined as follows
\begin{eqnarray}
&& D_\mu\Phi = \partial_\mu\Phi - ie[A_\mu,\Phi]_*\nonumber\\
&& F_{\mu\nu} = \partial_\mu A_\nu - \partial_\nu A_\mu
-ie[A_\mu,A_\nu]_*
\end{eqnarray}
The star product $*$ in (\ref{uno}) is defined as usual
\begin{equation}
A(x)*B(x) = \left. \exp\left( \frac{i}{2} \theta_{\mu
\nu}\partial^\mu_x
\partial^\nu_y \right)A (x)B(y)
\right\vert_{y=x}
 \label{moyal}
\end{equation}
and provides a simple way to handle field theories in noncommutative
space as that defined in (\ref{nc}). We have defined
\begin{equation}
[A(x),B(x)]_* = A(x)*B(x)- B(x)*A(x)
\end{equation}
As in eq.({\ref{nc}) we shall take $\theta_{0i} = 0$. This ensures a
well-defined Hamiltonian and unitarity at the quantum level.
Moreover, under such conditions, we shall see that the Noether
theorem can be naturally extended to the noncommutative case and
conserved charges can be derived avoiding the problems implied by
the infinite number of time derivatives that introduces $\theta_{0i}
\ne 0$.

Let us briefly recall at this point the way in which the $*$-product
induces charge violation. This can be easily seen just by analyzing
the pure gauge action. Under charge conjugation
 gauge fields change according to
\begin{equation}
A_\mu = A_\mu^a t^a \stackrel{C}{\Longrightarrow} A_\mu^C = -
\overline  A_\mu = - A_\mu^a \overline t^a \label{change}
\end{equation}
where $\overline A_\mu$ represents the complex conjugate
representation. One can easily see  \cite{Sheikh1} that already the
commutator  $[A_\mu^0,A_\nu^0]$ entering in the $F_{\mu\nu}^0$
component of the field strength does not change sign under charge
conjugation while the corresponding derivative terms do. There are
also mismatches concerning the $SU(2)$ components. Only if change
(\ref{change}) is accompanied by a change of sign in $\theta_{ij}$
these changes are compensated. Since we take $\theta_{ij}$ as fixed
parameters, we have
\begin{equation}
{\rm tr} (F_{\mu\nu} F_{\mu\nu}) \ne {\rm tr} (F^C_{\mu\nu}
F^C_{\mu\nu})
\end{equation}
and then the action (\ref{uno}) is not invariant under charge
conjugation. In contrast, since under parity transformation one has
\begin{equation}
A_\mu   \stackrel{P}{\Longrightarrow} A_\mu^P = \left
\{{}^{{}^{{}{\textstyle{~\;A_{0}}}}}_{{}_{{}{\textstyle{-A_{i}}}}}
\right.
\end{equation}
it is easy to see that these transformations together with those in
coordinates ($x^i\stackrel{P}{\Longrightarrow} - x^i$) leave   the
action (\ref{uno}) invariant. Regarding time inversions, one can see
that the action changes and such change,   compensates the one
produced by charge conjugation. In summary,  CP is violated but the
theory is CPT invariant.

We shall now look for noncommutative dyon solutions which, being
static, can be found by searching the minima of the energy, defined
as
\begin{equation}
E =  {\rm tr} \int d^3x  \left( E_i*E_i + B_i*B_i + D_i\Phi * D_i
\Phi + D_0\Phi* D_0\Phi \right) \label{energy}
\end{equation}
where we have written
\begin{equation}
E_i = - F_{0i} \; , \;\;\;  B_i = - \frac{1}{2} \varepsilon_{ijk}
F^{jk}
\end{equation}
Note that since we are working in the BPS limit, the vacuum
expectation value is no longer determined by the Lagrangian but
imposed as a boundary condition on the Higgs field
\be {\rm tr} \Phi_{vac}^2 =  \frac{v_0^2}{2} \label{condi} \ee
 As in ordinary space\cite{Coleman},  Eq.(\ref{energy})  can be written in the
  form
 \begin{eqnarray}
E &=& {\rm tr} \int d^3x \left(
 \left(E_i - \sin \alpha D_i\Phi\right) *
\left(E_i - \sin \alpha D_i\Phi\right) \right.\nonumber\\
&& +  \left(B_i - \cos \alpha D_i\Phi\right) * \left(B_i - \cos
\alpha D_i\Phi\right)
 + D_0\Phi* D_0\Phi\nonumber\\
&&+ \left.   2 \sin \alpha E_i * D_i\Phi + 2\cos\alpha B_i * D_i\Phi
\right )
 \end{eqnarray}
Thus, one has a Bogomol'nyi bound on the energy
\begin{equation}
E \geq  v_0 \sin \alpha Q +  v_0 \cos \alpha M
\end{equation}
with $Q$ and $M$ the electric and magnetic charges defined as
\begin{eqnarray}
Q &=& \frac{2}{v_0}{\rm tr} \int d^3x E_i * D_i\Phi \nonumber\\
M &=& \frac{2}{v_0}{\rm tr} \int d^3x B_i * D_i\Phi \label{cargas}
\end{eqnarray}
The bound is saturated whenever the following BPS equations hold
\begin{eqnarray}
&& E_i =  \sin \alpha D_i \Phi \label{a}\\
&& B_i =  \cos \alpha D_i \Phi \label{b}\\
&& D_0\Phi = 0 \label{c}
\end{eqnarray}
Since we are looking for static configurations, eq.(\ref{c}) implies
\begin{equation}
[A_0, \Phi]_* = 0 \label{conmutador}
\end{equation}

In order to find an explicit dyon solution  we  consider below  an
expansion of fields $A_\mu$ and $\Phi$ in powers of
 the noncommutative parameter in
$\theta$  thus extending to the dyon case the aproach developed in
\cite{H1} and \cite{H2} where a purely magnetically charged solution
was found, to first and second order in $\theta$ respectively.

\section{Dyon solution as a  $\theta$-expansion}

The approach in refs.\cite{H1}-\cite{H2} for obtaining a
non-commutative monopole solution  starts from  the exact
Prasad-Sommerfield solutionin ordinary space  \cite{PS}  as giving
the zeroth-order of an expansion in powers of $\theta$ for the
monopole in noncommutative space. Plugging this expansion into the
BPS equations, one obtains the noncommutative solution order by
order in $\theta$. We shall follow this approach but including the
$A_0$ component of the gauge field so as to solve eqs.
(\ref{a})-(\ref{c}) and construct the dyon solution.

We then take as zeroth order approximation for the $SU(2)$
components the Prasad-Sommerfield dyon solution \cite{PS},
\begin{eqnarray}
\Phi^{a(0)} &=& \big(v_0\coth(v_0 e
 \cos(\alpha)r)-\frac{1}{ e r\cos(\alpha)}\big)\frac{x^{a}}{r}\nonumber\\
A_i^{a(0)} &=& \cos(\alpha)\big(\frac{1}{e r \cos(\alpha)}-
\frac{v_0}{\sinh(v_0 e r\cos(\alpha))} \big) \epsilon_{aij}\frac{x^{j}}{r}\nonumber\\
A_0^{a(0)} &=& \sin(\alpha) \big( v_0 \coth(v_0 e
 \cos(\alpha)r)- \frac{1}{e r\cos(\alpha)}\big)\frac{x^{a}}{r}\nonumber\\
 \label{Ans0}
\end{eqnarray}
 Notice that
\begin{equation}
A_0^{a(0)} = \sin \alpha \Phi^{a(0)}
\label{prop}
\end{equation}
Concerning the $U(1)$ components, we take
\be \Phi^{0(0)} = 0\; , \;\;\;   A_i^{0(0)} = 0\; , \;\;\;
A_0^{0(0)} = 0\; , \;\;\;
\ee
In order to find the complete solution we write
\begin{eqnarray}
\Phi &=& (\Phi^{a (0)} +   \tilde \Phi^{a(1)} +
\tilde \Phi^{a (2)})t_{a}\nonumber\\
 &+&  (\Phi^{0 (0)} + \tilde \Phi^{0(1)}+
 \tilde \Phi^{0 2)})t_{0} +
 \mathcal{O}(\theta^{3})\nonumber\\
A_i &=& ( A_i^{a(0)} + \tilde  A_i^{a(1)}+
\tilde A_i^{a(2)}) t_{a}\nonumber\\
  &+&  (A_i^{0(0)} +\tilde  A_i^{0(1)} + \tilde A_i^{0(2)}) t_{0}+
  \mathcal{O}(\theta^{3})\nonumber\\
A_0 &=& (A_0^{a (0)} + \tilde A_0^{a (1)} +
\tilde A_0^{a (2)})t_{a}\nonumber\\
 &+&  (A_0^{0 (0)} + \tilde A_0^{0 (1)}+ \tilde A_0^{0 (2)})t_{0} +
 \mathcal{O}(\theta^{3}) \label{solut}
\end{eqnarray}
\underline{}To first order in $\theta$, an ansatz for the gauge
potencial  and Higgs field components on U(1), that obeys covariance
under the $SO(3)$ rotation corresponding to the diagonal  subgroup
of $SO(3)_{gauge} \times SO(3)_{space}$ is
\begin{eqnarray}
\tilde A_{i}^{0(1)} &=& \theta_{ij} x_{j} A(r) +  \varepsilon_{ijk}
\theta_{jk} C(r) + x_{i} \varepsilon_{jkl}
\theta_{jk} x_{l} D(r)\nonumber\\
\tilde A_0^{0(1)} &=& \theta_{ij} \varepsilon_{ijk} x_{k} K(r)\nonumber\\
\tilde \Phi^{0(1)} &=& \theta_{ij} \varepsilon_{ijk} x_{k} B(r)
\label{Ans1}
 \end{eqnarray}
where $A(r),B(r),C(r), D(r)$ and $K(r)$ are radial functions to be
determined. The component on $SU(2)$ to first order in $\theta$ of
the Bogomol'nyi equation is not going to be analyze, due to it have
a solution that is pure gauge.

Using ansatz\ae ~(\ref{Ans0}) and (\ref{Ans1}) one can easily show
that Bogomol'nyi eq.(\ref{conmutador}) implies, at this order in
$\theta$, that
\begin{equation}
K(r) = \sin \alpha B(r)
\end{equation}
 Extending the symmetric ansatz\ae
(\ref{Ans0})-(\ref{Ans1})
 order by order in $\theta$   one can prove  that this
 kind of relation remains valid so
 that one can conclude that, to all orders in $\theta$, one has
\begin{equation}
A_0 = \sin \alpha \Phi
\label{relation}
\end{equation}
A similar relation was proposed in \cite{GN} as an asatz  within Nahm's
approach to the construction of monopole solutions. Here, it was
derived as a result of taking the original dyon solution
 in ordinary space as
the zeroth order in a $\theta$-expansion.

Identity (\ref{relation}) implies that the BPS equation (\ref{a}) is
automatically satisfied and then
 the BPS system (\ref{a})-(\ref{b}) reduces to
\begin{equation}
  \frac{1}{2} \varepsilon_{ijk} F_{jk} = - \cos \alpha D_i \Phi \label{unica}
\end{equation}
Except for the factor $\cos \alpha$, this is nothing but the pure
noncommutative monopole equation. We can then use the pure monopole
solutions constructed in \cite{H1}-\cite{H2}   after making the
appropriate rescaling. We find
\begin{eqnarray}
\tilde A_i^{(n)}(x^i) &=& \cos^{2n+1}\alpha A_i^{(n)}(\cos \alpha
x^i) \, , \;\;\; n=1,2, \ldots
\nonumber\\
\tilde \Phi(x^i)^{(n)} &=& \cos^{2n}\alpha  \Phi^{(n)}(\cos \alpha
x^i) \, , \;\;\; n=1,2, \ldots
\end{eqnarray}
where $A_i^{(n)}$ and $ \Phi^{(n)}$ are the pure monopole solutions
found in \cite{H2}. Then, using eq.(\ref{relation}) $\tilde
A_0^{(n)}(x^i)$ can be constructed. We give their explicit form, to
order $\theta^2$, in an Appendix.

An important property of the solution we found is that, having
started at zeroth order  with $ \Phi^{0(0)} = 0$, one finds that
this condition holds to all orders in $\theta$. It should be noticed
that, as we already pointed, since we are working in the
Prasad-Sommerfield limit, the vacuum expectation vacuum is no longer
determined by the Lagrangian but imposed as a boundary condition on
the Higgs field, eq.(\ref{condi}). The vanishing of the $U(1)$
component of the Higgs field solution then implies that
(\ref{condi}) should be guaranteed by the $SU(2)$ components \be
(\Phi_{vac}^a)^2 = v_0^2 \; , \;\;\; \;\;\; \;\;\; \Phi_{vac}^{0} =
0 \label{conditionv} \ee

 We are now ready
to compute the electric and magnetic charges of the dyon solution.
Expanding the r.h.s of eqs.(\ref{cargas})
  we have
\begin{eqnarray}
  && Q = \frac{2}{v_0} {\rm tr} \int_{S^{2}_{\infty}} dS^{i} (\Ei \Phi +
  \frac{i}{2}\theta^{l m} \partial_{l}E_{i} \partial_{m} \Phi -
  \frac{1}{4}\theta^{p q}\theta^{l m}\partial_{p}\partial_{l}E_{i}
  \partial_{q}\partial_{m} \Phi)+\mathcal{O}(\theta^{3})\nonumber \\
&&  M = - \frac{2}{v_0} {\rm tr} \int_{S^{2}_{\infty}} dS^{i}
(\frac{1}{2}\eijk F_{jk} \Phi+ \frac{i \eijk}{4}\theta^{l m}
\partial_{l}F_{jk}
\partial_{m} \Phi \nonumber
\\&&
 - \frac{\eijk}{4}\theta^{p q}\theta^{l m}\partial_{p}\partial_{l}F_{jk}
\partial_{q}\partial_{m} \Phi)+\mathcal{O}(\theta^{3})
\end{eqnarray}
In computing these charges we shall need to know
 the asymptotic behavior of solutions (\ref{solut}), order
 by order in $\theta$.  Starting
from the behavior to order zero
\begin{eqnarray}
\label{DionASint}
  \Phi^{a(0)} &=& v_0 + \mathcal{O}(e^{-v_0r})\\
  A^{a(0)}_{i} &=& \frac{1}{e r} + \mathcal{O}(1/r^{2})+ \mathcal{O}(e^{-v_0r})\\
    A_0^{a(0)} &=& v_0 \sin \alpha  + \mathcal{O}(e^{-v_0r})\\
  \end{eqnarray}
  one has, to order $n$ ($n= 1,2, \ldots$)
  \begin{eqnarray}
 \Phi^{(n)} &=& \frac{f_0^{(n)}}{r^{n+1}} + \mathcal{O}(1/r^{n+2})+ \mathcal{O}(e^{-v_0r})\\
  A_{i}^{(n)} &=& \frac{f_i^{(n)}}{r^{n+1}} + \mathcal{O}(1/r^{n+2})+
  \mathcal{O}(e^{-v_0r})\\
    A_0^{(n)}  &=& \frac{f_0^{(n)}\sin \alpha}{r^{n+1}} +
    \mathcal{O}(1/r^{n+2})+ \mathcal{O}(e^{-v_0r})
    \label{asymn}
\end{eqnarray}
with $f_0^{(n)}$ and $f_i^{(n)}$ constants. From this behavior one
can see that $n>0$ orders  in $\theta$ do not contribute to the
charges that thus coincide with those in ordinary space,
\begin{eqnarray}
  Q &=& \frac{1}{ v_0}
  \int_{S^{2}_{\infty}} dS^{i} (\partial_{i}A^{a(0)}_{0}+ e
   \epsilon_{abc}A^{b(0)}_{i}A^{c(0)}_{0}) \Phi^{a(0)} =
\frac{4\pi}{e}    \tan \alpha
   \\
M & = & - \frac{1}{  v_0} \int_{S^{2}_{\infty}} dS^{i}  \eijk
(\partial_{j}A^{a(0)}_{k} +\frac{e}{2}
  \epsilon_{abc}A^{b(0)}_{j}A^{c(0)}_{k})  \Phi^{a(0)} = \frac{4\pi}{e}
\end{eqnarray}
It could be argued that this  coincidence is just a consequence  of
having constructed the solution  as a power expansion in $\theta$,
which has dimensions of $[{\rm length}]^{2}$, so that higher orders
in $\theta$ would necessarily imply, asymptotically,  higher powers
in $1/r$. There is however another dimensional parameter, $v_0^2$
($[v_0^2] = [1/{\rm length}]^{2}$) so that (\ref{asymn}) is the
result of
 the structure of BPS equations and the boundary conditions and
 not just a dimensional question.

Note that the following relation between charges holds
\begin{equation}\label{dion}
 Q = M  \tan\alpha\; ,
\end{equation}
the same one satisfied by  dyon electric and magnetic charges in
ordinary space. We then conclude that CP violation induced by
noncommutativity does not change, at least at the classical level,
the charge of the dyon: no trace of $\theta_{ij}$ appears in $Q$.

In order to analyze  charge quantization at the quantum level, we
can use the Noether approach and canonically proceed
 as originally done
in
\cite{Witten} regarding the unbroken symmetry which leaves the
Higgs vacuum invariant and is associated to the electric charge.

\section{The Noether theorem and the dyon charge shift in the
noncommutative case}

In order to analyze the Noether charge we shall not just consider
action (\ref{uno}) but one including the noncommutative
version of a $\vartheta$-term,
\begin{equation}
S = {\rm tr} \int d^4 x \left(-\frac{1}{2} F_{\mu\nu} * F^{\mu\nu} +
 D_\mu \Phi * D^\mu\Phi + \vartheta \frac{e^2}{16\pi^2}
F_{\mu\nu} * \tilde F^{\mu\nu}  \right)
 \label{unotheta}
\end{equation}
In this way, we shall be able to test possible modifications of the
dyon charge both because noncommutativity and because the
$\vartheta$ term. (Of course the case of action (\ref{uno}) can be
analyzed just by putting $\vartheta = 0$). As in ordinary space, the
noncommutative version of the $\vartheta$ term  can be written
as a surface
term \cite{Furuuchi} and hence the equations of motion for action
(\ref{unotheta}) and its dyon
solutions remain the same for all values of $\vartheta$.

After symmetry breaking through condition (\ref{conditionv}), the
unbroken symmetry is related to rotations $\Lambda$ in the direction
of $\Phi$, which at infinity satisfy \be \Lambda_\infty =
\frac{1}{
v_0} \Phi^a_{vac} t^a
 \ee
  clearly leaving
 the Higgs vacuum invariant. We thus consider such gauge transformations
along the Higgs field direction \be \Lambda(x) = \frac{1}{ v_0}
\Phi^a  t^a \epsilon(x) \ee
with $\epsilon(\infty) = 1$. With this, one has
 \begin{eqnarray}
 \delta_\Lambda  \Phi &=& 0\nonumber\\
 \delta_\Lambda A_\mu &=& \frac{1}{ ev_0} D_\mu\Phi
 \end{eqnarray}

When one looks for field tranformations living the action $S$
unchanged in order to find a conserved current in the noncommutative
case, one has to take into account $*$-commutators that, once
integrated, give a vanishing contribution, as first noticed for a
scalar theory in \cite{MS}. To see this, let us consider a  gauge
transformation $\Lambda$ with infinitesimal parameter $\epsilon(x)$.
Being the action $S$ invariant in the case $\epsilon = constant$,
one can write for the local case
\be \delta_{\Lambda(x)}S = - \int d^4 x J^\mu[\Phi(x),A_\mu(x)]
 \partial_\mu \epsilon(x)
\ee
or, after integrating by parts
\be \delta_{\Lambda(x)}S =  \int d^4 x \partial_\mu
J^\mu[\Phi(x),A_\mu(x)]  \epsilon(x) \ee Invariance under local
transformations implies
 $\delta_{\Lambda(x)} S = 0$ or
\be
 \int d^4 x \partial_\mu J^\mu[\Phi(x),A_\mu(x)] \epsilon(x) = 0
\label{noe} \ee
 It is at this point that the Noether's procedure
deviates from the usual one in ordinary space. Indeed, in the
noncommutative case  the most one can infer from eq.(\ref{noe}) is
that
\be \left( \partial_\mu J^\mu \right) = {\rm tr} \left( [O,P]_* +
[B,C]_* *B + \ldots \right)\label{que} \ee
for some proper functionals $O,P,B,C, \dots$ since,
once integrated over space-time, the r.h.s.  in (\ref{que}) vanishes due
to the $*$-product cyclic
properties under integration.

In order to find a conserved charge one has to integrate (\ref{que})
over 3-space. Now, since we have taken $\theta_{0i} = 0$ also in
this case the integral of commutators vanishes, \be {\rm tr} \int
d^3x \left( [O,P]_* + [B,C]_* *B + \ldots \right) = 0 \ee
so that, after integrating (\ref{que})
 one ends with a
conserved charge of the form
\be
N = \int d^3x J^0
\ee
Use of the equations of motion allows to explicitly
find
\be J^0 = -\frac{1}{ e v_0}{\rm tr} \left( 2 F^{0i}* D_i\Phi -
\frac{\vartheta  e^2}{4\pi^2} \tilde F^{0i} * D_i\Phi  \right) \ee
This implying, in the present case
\be N = \frac{1}{ e v_0}{\rm tr} \int d^3x \left(2 F_{0i}* D_i\Phi -
\frac{\vartheta e^2}{8\pi^2}
 \varepsilon_{ikj}F_{jk}* D_i\Phi
 \right) = -\frac{1}{e} Q
+ \frac{\vartheta e}{8\pi^2} M
\label{N}
\ee
At large distances, where ${\rm tr} \Phi^2 = v_0^2/2$ the operator
$\exp(2\pi i N)$ is a $2\pi$ rotation about the direction of $\Phi$
(elsewhere the rotation angle is $2\pi |\Phi(x)|/  v_0$ but by Gauss
law if the gauge transformation is 1 at infinity it leaves the
physical states invariant) . Then we have \be \exp(2\pi i N) = 1 \ee
and thus the eigenvalues of $N$ have to be quantized in integer
units $n$. If we call $q_e$ and $q_g$ the eigenvalues of the
electric and magnetic charge operators, one gets from (\ref{N}) \be
q_e = \left (n e + \frac{\vartheta e^2}{8\pi^2}q_g\right)
\label{final} \ee
That is, we have obtained for the nonconmmutative dyon the same formula
that holds for the case of ordinary space, eq.(\ref{effect}).

In summary, we have constructed an explicit noncommutative dyon
solution showing that the  relation (\ref{dion}) between classical
electric and magnetic charge also holds in noncommutative space.
Moreover, after extending the Noether approach to the case of a
noncommutative gauge theory, we have proven that the $\theta_{ij}$
dependent CP violation introduced by the commutation rule (\ref{nc})
does not change the Witten effect formula; indeed, the dyon's charge
shift is $\theta_{ij}$-independent  for constant parameters
$\theta_{ij}$. In this respect, it should be interesting to consider
other type of noncommutativity and in particular, to investigate the
case of the dyon in the fuzzy sphere along the lines developed in
ref.\cite{Moreno} where monopole solutions were constructed for the
case in which $\theta_{ij} = \theta r \varepsilon_{ijk}x_k$ since in
that case the coordinate dependence of $\theta_{ij}$ may introduce
definite changes in (\ref{final}). We hope to discuss this issue in
the future.

\vspace{1 cm}

\noindent\underline{Acknowledgments}

\noindent We would like to thank the Sociedad Cientifica Argentina
for hospitality. We are grateful to acknowledge G.~Giribet and
G.~Lozano for interesting discussions and comments.
 This work is partially supported by CONICET (PIP6160), ANPCyT (PICT 20204),
 UNLP and CICBA  grants.

\section*{Appendix}
Given the BPS equation
\begin{equation}
\frac{1}{2}\varepsilon_{ijk}F_{jk} + D_{i}\Phi=0
\end{equation}
the monopole solution,  taking as zeroth order in $\theta$
the Prasad-Sommerfield solution (\ref{Ans0}), is, up to order $\theta^2$,
\begin{eqnarray}
 A_{i}^{a(1)} &=& 0\nonumber\\
 \Phi^{a(1)} &=& 0\nonumber\\
 A_{i}^{0(1)} & = &\theta_{ij} x_{j} \frac{1}{4r^{2}} W(W+2F)\nonumber\\
 \Phi^{0(1)} & =& 0
\nonumber\\
  A_{i}^{a(2)} & = & \frac{1}{r^{5}} \bigg[ a_{1}(r) \theta^{2} \epsilon_{aij} \hat{x}_{j}
    + a_{2}(r)(\theta \hat{x}) \epsilon_{aij} \theta_{j} +
   a_{3}(r) \epsilon_{ajk} \theta_{i} \theta_{j} \hat{x}_{k}
     \nonumber\\
     &&+  a_{4}(r) (\theta \hat{x})^{2} \epsilon_{aij} \hat{x}_{j}
     + a_{5}(r) (\theta \hat{x}) \epsilon_{ajk}
  \hat{x}_{i} \theta_{j} \hat{x}_{k} \bigg]\nonumber\\
  \Phi^{a(2)} &=& \frac{1}{r^{5}} \bigg[ \phi_{1}(r) (\theta
\hat{x}) \theta_{a} + \phi_{2}(r)
  \theta^{2} \hat{x}_{a} + \phi_{3}(r)(\theta \hat{x})^{2} \hat{x}_{a}
  \bigg]\nonumber\\
  A_{i}^{0(2)} &=& 0\nonumber\\
    \Phi^{0(2)} &=& 0
\end{eqnarray}
Here
\begin{eqnarray}\label{Hatasolucion}
    \phi_{1}(r) \!\!&=&\!\! -\frac{1}{4}r F + \frac{1}{4} r^{2} F^{2} -
    \frac{1}{8} r^{3} F^{3} +
  \frac{1}{4}r F (1- r W )^{2}\nonumber\\
  \phi_{2}(r) \!\!&=&\!\! \frac{1}{8} - \frac{3}{8} r F + \frac{1}{8}r^{2} F^{2} -
  \frac{1}{4} (1 - r W)^{2}
  + \frac{3}{8}r F (1- r W)^{2} + \frac{1}{8} (1- r W)^{4}\nonumber\\
  \phi_{3}(r) \!\!&=&\!\! -\frac{1}{8} + \frac{7}{8} r F - \frac{5}{8} r^{2} F^{2} +
  \frac{1}{4} (1 -r W)^{2}
  + \frac{1}{8} r^{3} F^{3} - \frac{7}{8} r F (1 - r W)^{2}\nonumber\\
  && \!\! - \frac{1}{8} r F^{2} (1- r W)^{2} -\frac{1}{8} (1 - r W)^{4}
\end{eqnarray}
and
\begin{eqnarray}
  a_{1}(r) & = & - \frac{1}{8} + \frac{1}{2} r F - \frac{1}{8} (1 - r W) -
  \frac{1}{2} r F (1 - r W)
  - \frac{1}{4} r^{2} F^{2} (1 -r W)\nonumber
\\
  & & + \frac{5}{8} (1 - r W)^{2}+ \frac{1}{4}r F (1 -r W)^{2} - \frac{3}{8} (1 - r W)^{3}
  - \frac{1}{4}r F (1 - r F)^{3}\nonumber\\
  a_{2}(r) & = & \frac{1}{4} + \frac{1}{2} r F - \frac{3}{8}r^{2} F^{2} - \frac{3}{4}(1- r W)
  -\frac{1}{2} r F (1 - r W) \nonumber\\
  & & +
   \frac{1}{8} r^{2} F^{2} (1 - r W)
   + \frac{3}{4}(1 - r W)^{2} - \frac{1}{4}(1 - r W)^{3}
    \end{eqnarray}
  \begin{eqnarray}
  a_{3}(r) &= & -\frac{1}{8}- \frac{1}{4}r F- \frac{1}{8}r^{2}F^{2} + \frac{3}{8}(1 -r W)
  + \frac{1}{2}r F (1 - r W) \nonumber\\
  & & +\frac{1}{8} r^{2} F^{2} (1 - r W) - \frac{3}{8} (1 - r W)^{2} -
  \frac{1}{4} r F (1 - r W)^{2}\nonumber\\
  &&  + \frac{1}{8} (1 - r W)^{3}\nonumber\\
  a_{4}(r) &= & -\frac{1}{4} - \frac{3}{2} r F + \frac{1}{2} + \frac{1}{2} r^{2} F^{2}
  + \frac{5}{4} (1 - r W) + \frac{3}{2} r F (1- r W)\nonumber \\
  & & + \frac{1}{4} r^{2} F^{2} (1 -r W) - \frac{7}{4}r^{2} F^{2} -
  \frac{1}{4} r F (1 - r W)^{2}\nonumber\\
  & & + \frac{3}{4}(1 -r W)^{3} + \frac{1}{4} r F ( 1- r W)^{3}\nonumber\\
  a_{5}(r) &= &0
  \end{eqnarray}
with
\begin{eqnarray}\label{CommAnsatz2}
  F(r) & =& v_0 \coth(v_0 e r)-\frac{1}{e r}\nonumber \\
  W(r) &=& \frac{1}{e r}-\frac{v_0}{\sinh(v_0 e r)}
\end{eqnarray}
and
\begin{eqnarray}
   \theta_{i} & \equiv & (1/2) \eijk \theta^{jk}\nonumber\\
 \theta^{2} & \equiv & \theta_{i} \theta_{i}\label{const}\\
 (\theta \hat{x}) & \equiv & \theta_{i} \hat{x}_{i}\nonumber
  \end{eqnarray}

\end{document}